\documentclass[pra,aps,showpacs,showkeys,preprint]{revtex4}
\usepackage{graphicx}
\usepackage{amsmath}
\begin{document}
\title{\bf The Minimal Length and the Quantum Partition Functions}
\author{M. Abbasiyan-Motlaq}
\author{Pouria Pedram}
\email[Electronic address: ]{p.pedram@srbiau.ac.ir}
\affiliation{Department of Physics, Science and Research Branch,
Islamic Azad University, Tehran, Iran}
\date{\today}

\begin{abstract}
We study the thermodynamics of various physical systems in the
framework of the Generalized Uncertainty Principle that implies a
minimal length uncertainty proportional to the Planck length. We
present a general scheme to analytically calculate the quantum
partition function of the physical systems to first order of the
deformation parameter based on the behavior of the modified energy
spectrum and compare our results with the classical approach. Also,
we find the modified internal energy and heat capacity of the
systems for the anti-Snyder framework.
\end{abstract}


 \maketitle

\section{Introduction}
String theory, loop quantum gravity, quantum geometry, and
black-hole physics as the candidates of quantum gravity all imply
the existence of a minimal measurable length of the order of the
Planck length $\ell_{\mathrm{P}} =
\sqrt{\frac{G\hbar}{c^{3}}}\approx 10^{-35}$m, where $G$ is Newton's
gravitational constant \cite{felder,1,2,3,4,5,6,7,8}. To incorporate
the idea of the minimal length with the laws of quantum mechanics,
we need to modify the Heisenberg uncertainty principle to the
Generalized (Gravitational) Uncertainty Principle (GUP). Based on
the Heisenberg uncertainty principle, the minimal measurable length
$\Delta X$ can be made arbitrarily small by letting $\Delta P$ grow
correspondingly, but it fails near Planck energy scale where the
corresponding Schwarzschild radius and Compton wavelength become
approximately equal to the Planck length.

In the GUP framework, because of the modification of the commutation
relations between the momentum and the position operators, all the
Hamiltonians in quantum mechanics will be modified which results in
the generalized Schr\"odinger equation in the quantum domain.
Although the generalized Schr\"odinger equation is a high order
differential equation and it is not easy to solve it, various
problems such as the harmonic oscillator, the hydrogen atom, the
Casimir effect, gravitational quantum well, and particles scattering
have been investigated exactly or approximately in the GUP scenario
\cite{9,10,11,12,13,14,15,16,17,17-2,18,19,20,21,22,23,24,25}.

Deliberation of many topics in theoretical physics are based on
statistical approaches. In the GUP framework Fityo developed a
classical method of the partition function evaluation based on
modification of elementary cell of space volume according to
modification of the commutation relation and investigated the
thermodynamical properties of some physical systems \cite{27}. The
internal energy and the heat capacity of various physical systems
are also calculated in Refs.~\cite{26,27,32}. Moreover, the effects
of the minimal length on quantum partition function, internal energy
and heat capacity using the exact treatment of the generalized
Schr\"odinger equation is studied numerically in Ref.~\cite{17}.

In this paper, we investigate the thermodynamical properties of the
physical systems up to the first order of the GUP parameter. We
present a general scheme to analytically calculate the modified
partition function in term of ordinary partition function and the
behavior of the modified energy spectrum of the physical systems.
Here, we obtain the internal energy and heat capacity of
one-dimensional ideal gas, harmonic oscillator, and Landau levels
and compare our results with those obtained by the semiclassical
approach. Also, we study these problems in anti-Snyder scenario
based on our general scheme.

\section{The Generalized Uncertainty Principle}
In one dimension the modified commutation relation according to
Kempf \emph{et al.} is \cite{10}
\begin{equation}\label{001}
[X,P]=i \hbar (1+ \beta P^2 ),
\end{equation}
that leads to the following generalized uncertainty relation
\begin{equation}\label{002}
\Delta X  \Delta P \geq \frac{ \hbar }{2} ( 1+ \beta  (\Delta P)^2 +
\zeta ),
\end{equation}
and a nonzero minimal measurable length $ ( \Delta X ) _{min} =
\hbar \sqrt{ \beta } = \sqrt{ \beta _{0} } \ell_{\mathrm{P}}$. Here
$\zeta$ is a constant and $ \beta = \beta _{0} / (c ^{2}
M^{2}_{{\mathrm{P}}})^2 $ is the GUP parameter, $ M_{{\mathrm{P}}} =
\sqrt{\hbar c / G}$ is the Planck mass, and $ \beta _{0} $ is a
dimensionless parameter of the order of unity. Following
Ref.~\cite{10} we can write $ X $ and $ P $ in the momentum space
representation as
\begin{eqnarray}\label{003}
X\psi(p)& =&i \hbar ( 1 + \beta p^{2} ) \frac{ \partial \psi(p)}{
\partial p },\\
P\psi(p)&=&p\,\psi(p), \label{004}
\end{eqnarray}
where $ X $ and $ P $ are symmetric operators and exactly satisfy
the commutation relation (\ref{001}). However, in general,
expressing the position operator as a composition of ordinary
position and momentum operators would complicate the resulting
differential equation. To overcome this problem and since the above
representation is not unique, we can define the following
quasi-position representation \cite{17}
\begin{eqnarray}\label{007}
X\psi(\xi)&=&\xi\,\psi(\xi), \\
P\psi(\xi)&=&\frac{\tan (-i\hbar \sqrt{\beta} \, \partial_\xi )
}{\sqrt{\beta}}\psi(\xi),\label{008}
\end{eqnarray}
To first order of GUP parameter, this representation can be written
as \cite{28}
\begin{eqnarray}\label{9}
 X\psi(\xi)&=&\xi\psi(\xi), \\
P\psi(\xi)&=&-i\hbar\left(1-\frac{1}{3} \, \beta \, \hbar^2 \frac{
\partial^2 }{ \partial \xi^2 }\right)\frac{
\partial \psi(\xi)}{ \partial \xi }.\label{10}
\end{eqnarray}
Note that this representation is on the space of quasi-position wave
functions not on the space of position wave functions. The
quasi-position representation is firstly introduced by Kempf
\emph{et al.} in their seminal paper about twenty years ago
\cite{10} and extensively used by others during the past two
decades. In the presence of a minimal uncertainty in position, the
position operator is only symmetric and is no longer essentially
self-adjoint. Nevertheless, the symmetricity implies that all
expectation values are real. However, since the Heisenberg algebra
representation contains no position eigenstates $|X\rangle$, there
is no Hilbert space representation on position wave functions
$\langle X|\psi\rangle$ in this algebra. Indeed, because the formal
position eigenfunctions do not obey the minimal length uncertainty,
they are not physical states.

Notice that, it is possible to project $|\psi\rangle$ on maximally
localized states $|\psi^{ml}_\xi\rangle$ to find the probability
amplitude for the maximally localized particle around the position
$\xi$, i.e., $\xi=\langle \psi^{ml}_\xi| X|\psi^{ml}_\xi\rangle$.
The maximal localization states are proper physical states and we
can use them to define a `quasi-position' representation. Although
this representation does not diagonalize $X$, it has a direct
interpretation in terms of position measurements. Therefore, the
position and momentum operators $X$, $P$ can be written in terms of
the multiplication and differentiation operators $\xi$, $p=-i\hbar
\frac{d}{d\xi}$ that satisfy the ordinary commutation relations of
quantum mechanics, namely, $[\xi,p]=i\hbar$. In Ref.~\cite{17-2},
the maximally localized states are obtained in quasi-position
representation given by Eqs.~(\ref{007},\ref{008}).

In fact, we do not work with the same space of physical states with
the same properties as we deal in ordinary quantum mechanics. For
quasi-position wave functions and only for $\beta\rightarrow0$, the
scalar products (\ref{007},\ref{008}) agree with the ordinary scalar
products on position space. Moreover, the quasi-position
representation does not diagonalize the position operator in any
domain of the symmetric operators $P^2$ and $X^2$. The direct
physical interpretation of the quasi-position representation is the
main preference of this representation. $\psi(\xi)$ denotes the
probability amplitude for finding the particle which is maximally
localized around the position $\xi$ and with the uncertainty
$(\Delta X)_{|\psi^{ml}_\xi\rangle}=\hbar\sqrt{\beta}$ (see
Refs.~\cite{17,17-2} for more details).

To study the effects of GUP on the quantum mechanical systems,
consider the Hamiltonian
\begin{equation}\label{11}
H=\frac{P^{2}}{2m} +V(X),
\end{equation}
which using Eqs.~(\ref{007}) and (\ref{008}) can be expressed as
\cite{17}
\begin{equation} \label{12}
\qquad H=\frac{\tan ^{2} (\sqrt{\beta} \, p )}{2{\beta} \, m}
+V(\xi),
\end{equation}
with the following perturbative form
\begin{equation}\label{13}
H=H_{0} +\sum _{n=3} ^{\infty} \frac{ (-1)^{n-1} \, 2^{2n} \,
(2^{2n}-1) \, (2n-1) \, B_{2n}}{2m(2n)!} \beta^{n-2} \, p^{2(n-1)},
\end{equation}
where $ H_{0} = \frac{P^{2}}{2m} + V(\xi) $ and $ B_{n} $ is the nth
Bernouli number. Now in quasi-position space the Schr\"odinger
equation generalizes to \cite{17}
\begin{equation}\label{14}
-\frac{\hbar^{2}}{2m}\frac{\partial^{2}\psi(\xi)}{\partial
\xi^{2}}+\sum_{n=3}^{\infty} \alpha _{n} \,  \hbar^{2(n-1)} \,
\beta^{(n-2)}\frac{\partial^{2(n-1)}\psi(\xi)}{\partial
\xi^{2(n-1)}}+V(\xi) \, \psi(\xi)=E \, \psi(\xi),
\end{equation}
where $ \alpha _{n}=2 ^{2n} (2^{2n}-1)(2n-1) B _{2n} /2m (2n) !. $
As we shall see, the presence of the additional terms in the
modified Hamiltonian Eq.~(\ref{13}) leads to a positive shift in the
energy spectrum of a particle. This modified differential equation
is solved for the free particle, delta function potential, particle
in a box and  harmonic oscillator and their exact solutions are
obtained \cite{17}.

\section{Quantum statistical mechanics}
All physical systems can be studied in quantum statistical mechanics
by the partition function
\begin{equation}\label{15}
Z = \sum _{n} e^{-b E_{n}},
\end{equation}
where $ E _{n} $ is energy spectrum, $ b = \frac{1}{ k_{B} T } $, $
k_{B} $ is Boltzman's constant, and $ T $ is the temperature. Since
in the GUP scenario the generalized Schr\"odinger equation becomes a
high order differential equation, it is not an easy task to solve it
in general. However, for some specific problems this equation is
solved and the solutions are already obtained \cite{17,10,12,29,30}.
For instance, for a particle with mass $ m $ that is confined in an
infinite one-dimensional box with length $ L $, the corresponding
energy eigenvalues are \cite{17}
\begin{equation}\label{17}
E_{n} = \frac{ \tan ^{2} \sqrt{ 2m \, \beta \, E_{n} } }{ 2 m \beta
},
\end{equation}
where $ E_{n}^{0} = \frac{ n^{2} \pi ^{2} \hbar ^{2} }{ 2 m L^{2} }
$ are the eigenvalues of the ordinary Schr\"odinger equation. To
first order we have
\begin{equation}\label{18}
E_{n} = E_{n} ^{0} + \frac{4m \, \beta }{3} {E_{n} ^{0}}^{2}.
\end{equation}
Also for the Harmonic oscillator potential $ V(x) = \frac{1}{2} m
\omega ^{2} x^{2} $ we obtain \cite{10}
\begin{eqnarray}\label{19}
E_{n} &=& \hbar \, \omega \left( n + \frac{1}{2} \right)  \left(
\sqrt{1 + \frac{\gamma ^{2}}{4} } + \frac{ \gamma }{2} \right) +
\frac{1}{2} \, \hbar \, \omega \, \gamma \, n^{2},\\
&=&E_{n}^{0} + \frac{1}{2} \, m \, \beta \, { E_{n}^{0}
}^{2}+\frac{1}{8} \beta  m  \hbar ^{2}  \omega
^{2}+\mathcal{O}(\beta^2),\label{192}
\end{eqnarray}
where $ \gamma = m \hbar \omega \beta$ and $E_{n}^{0}=\hbar  \omega
\left( n + \frac{1}{2} \right)$. As it is shown in Ref.~\cite{33},
for the general polynomial potential in the form $ V(x)= | a|
x^{2(j+1)} +bx^{2(j+1)-1}+cx^{2(j+1)-2}+\ldots $ the approximate
energy spectrum for $ n\gg 1 $ is given by
\begin{equation}\label{i33}
E_{n} \simeq E_{n} ^{0} + \mathcal{O}(1) m\beta {E_{n} ^{0}}^{2} .
\end{equation}

Now to find the thermodynamical variables we need to compute the
partition function (\ref{15}). Suppose the energy levels of systems
have the form
\begin{equation}\label{21}
E_{n} = E_{n}^{0} + \varepsilon \, { E_{n}^{0} }^{2}+\Delta,
\end{equation}
then
\begin{equation}\label{22}
\qquad Z(b; \varepsilon ) = \sum _{n} \exp \left[ -b \left(
E_{n}^{0} + \varepsilon { E_{n}^{0} }^{2} + \Delta \right) \right].
\end{equation}
We now define $ Z^{0} ( b ) \equiv \sum \limits _{n} \exp \left( -b
E_{n}^{0} \right) $, so we have
\begin{equation}\label{23}
Z(b;\varepsilon ) =e^{-b\Delta} \sum _{m=0} ^{\infty} \frac{( -b \,
\varepsilon )^{m} }{ m ! } \frac{ \partial ^{2m} }{ \partial b^{2m}
} Z^{0} (b).
\end{equation}
For $ b \varepsilon \ll 1 $ or $ T \gg \frac{ \varepsilon }{k_{B}} $
and  up to the first order of $ \varepsilon $ the above equations
takes the form
\begin{equation}\label{24}
Z(b;\varepsilon ) =e^{-b\Delta}\left( Z^{0}(b) -b \varepsilon \frac{
\partial ^{2} }{
\partial b^{2} } Z^{0} (b)\right).
\end{equation}
For a system that consists of $ N $ particles we can write
\begin{equation}\label{25}
\qquad \ln Z_{N} ( b ; \varepsilon ) =  \ln Z_{N}^{0} (b)+N \ln
\left[1-b\, \varepsilon  \frac{\frac{\partial ^{2}}{\partial b^{2}}
Z^{0} (b)}{ Z^{0} (b)} \right]-b\Delta.
\end{equation}
Now, for $1 -b \, \varepsilon  \, \frac{ \frac{ \partial ^{2} }{
\partial b^{2} } Z^{0} (b) }{ Z^{0}(b) } > 0$
and using $ \ln (1-x) \simeq -x$ for small $x$ ($x\ll1$), we obtain
\begin{equation}\label{27}
\ln Z_{N} (b ; \varepsilon ) = \ln Z_{N}^{0} (b) - \varepsilon \, N
\, b \,  \frac{ \frac{ \partial ^{2} }{ \partial b^{2} } Z^{0} (b)
}{ Z^{0}(b) }-b\Delta.
\end{equation}
So the partition function to $\mathcal{O}(\varepsilon)$ is
\begin{equation}\label{28}
Z_{N}(b; \varepsilon )=Z_{N}^{0}(b) \,  \exp \left[ - \varepsilon \,
N \, b \, \frac{ \frac{\partial ^{2} }{ \partial b^{2} } Z^{0}(b) }{
Z^{0}(b) } -b\Delta\right].
\end{equation}
Also the modified internal energy $ E= -  \frac{ \partial }{
\partial b } \ln Z $ and the heat capacity $ C=\frac{
\partial E}{ \partial T} $, Eq.~(\ref{27}) are given by
\begin{eqnarray}\label{29}
E&=&E^{0}+ \varepsilon\frac{\partial }{\partial b}  \left[  N \, b
\, \frac{ \frac{\partial ^{2} }{\partial b^{2} } Z^{0}
(b)}{Z^{0}(b)}
\right]+\Delta+\mathcal{O}(\beta^{2}),\\
C&=&C^{0}+\varepsilon\frac{\partial }{\partial T}\frac{\partial
}{\partial b} \left[ N \, b \, \frac{\frac{\partial ^{2}}{\partial
b^{2}}Z^{0}(b) }{Z^{0}(b)} \right]+\mathcal{O}(\beta^{2}),\label{30}
\end{eqnarray}
where $ E^{0} $ and $ C^{0} $ are respectively the internal energy
and the heat capacity  in ordinary  thermodynamics framework.
Eqs.~(\ref {29}) and (\ref{30}) show that the shift in these
variables can be just calculated in terms of ordinary quantum
partition function $Z^{0}(b)$. Moreover, because of Eq.~(\ref{i33}),
this method can be applied to a wide range of problems.

It is worth to note that, we can also use these results for
anti-Snyder algebra \cite{18}
\begin{equation}\label{100}
[X,P]=i \hbar (1- \beta P^2 ),
\end{equation}
then by $\varepsilon \rightarrow-\varepsilon $ , Eqs.~(\ref{29}) and
(\ref{30}) become
\begin{eqnarray}\label{101}
E&=&E^{0}- \varepsilon\frac{\partial }{\partial b} \left[  N \, b \,
\frac{ \frac{\partial ^{2} }{\partial b^{2} } Z^{0} (b)}{Z^{0}(b)}
\right]+\Delta+\mathcal{O}(\beta^{2}),\\
 C&=&C^{0}-  \varepsilon\frac{\partial }{\partial
T}\frac{\partial }{\partial b} \left[ N \, b \, \frac{\frac{\partial
^{2}}{\partial b^{2}}Z^{0}(b) }{Z^{0}(b)} \right]+\mathcal{O}(\beta^{2}),\label{102}
\end{eqnarray}
where $ E $ and $ C $ are the internal energy and heat capacity of
the physical systems in anti-Snyder scenario, respectively.

\section{Applications}
In the following subsections we investigate the effects of GUP on
the one-dimensional ideal gas, harmonic oscillator and the Landau
levels.

\subsection{One-dimensional ideal gas}
Consider $ N $ noninteracting particles with mass $ m $ confined in
one-dimensional box with length $ L $. From Eqs.~(\ref{18}) and
(\ref{21}) we obtain $ \varepsilon = \frac{4}{3} m \beta $,
$\Delta=0$, and in ordinary statistical mechanics we have \cite{31}
\begin{equation}\label{31}
Z^{0} (b)=\frac{L (2 \pi \, m)^{1/2}}{h \, b^{1/2}}.
\end{equation}
Using Eqs.~(\ref{29}), (\ref{30}), and (\ref{31}) we obtain
\begin{eqnarray}\label{32}
 E&=&\frac{1}{2} Nk_{B}T - \beta  N  m  k_{B}^{2} T^{2}+\mathcal{O}(\beta^{2}), \\
 C&=&\frac{1}{2} Nk_{B}-2 \beta  N  m  k_{B}^{2}
T+\mathcal{O}(\beta^{2}),\label{33}
\end{eqnarray}
which are valid in the small temperature limit. Conditions
\begin{eqnarray}\label{shart}
1 -b \, \varepsilon  \, \frac{ \frac{ \partial ^{2} }{
\partial b^{2} } Z^{0} (b) }{ Z^{0}(b) } > 0, \hspace{2cm}b \, \varepsilon  \, \frac{ \frac{ \partial ^{2} }{
\partial b^{2} } Z^{0} (b) }{ Z^{0}(b) } \ll 1,
\end{eqnarray}
impose a restriction on temperature $ T $. In the case of
one-dimensional ideal gas this restriction is
\begin{equation}\label{34}
\frac{4 \beta m}{3k_{B}} \ll T < \frac{1}{\beta m k_{B}},
\end{equation}
where for an  electron with mass   $ m\sim10^{-30} kg$, we have $
10^{-9}K \ll T < 10^{54}K $. Note that, as the above equation
indicates, the temperature cannot be too small. From Eqs.~(\ref{32})
and (\ref{33}) it can be seen that to $\mathcal{O}(\beta)$ the
change in the internal energy and heat capacity of the system are
quadratic and linear functions of the temperature, respectively, and
at $T= \frac{1}{4\beta m k_{B}}$ the heat capacity goes to zero.
This result is also obtained in semiclassical approach of
Ref.~\cite{27}. By taking $ \beta \rightarrow - \beta $ in
Eqs.~(\ref{32}) and (\ref{33}), for anti-Snyder GUP and in the small
$T$ limit we have
\begin{eqnarray}\label{104}
 E&=&\frac{1}{2} Nk_{B}T + \beta  N  m  k_{B}^{2} T^{2}+\mathcal{O}(\beta^{2}),  \\
C& =&\frac{1}{2} Nk_{B}+2 \beta  N  m  k_{B}^{2}
T+\mathcal{O}(\beta^{2}),\label{105}
\end{eqnarray}
In contrast to Eqs.~(\ref{32}) and (\ref{33}) the above equations
indicate the increasing behavior of both internal energy and heat
capacity in this framework.

\subsection{Harmonic oscillator}
In ordinary quantum statistical mechanics, a  one-dimensional single
harmonic oscillator's partition function is given by
\begin{equation}\label{35}
Z^{0}(b)=\left[ 2 \sinh \left( \frac{b \hbar \omega}{2} \right)
\right]^{-1}.
\end{equation}
According to Eq.~(\ref{192}), $\varepsilon$ and $\Delta$ can be
written as
\begin{equation}\label{36}
\varepsilon = \frac{1}{2} m \beta, \hspace{2cm}\Delta=\frac{1}{8}
\beta m \hbar ^{2}  \omega ^{2},
\end{equation}
and consequently, Eqs.~(\ref{29}) and (\ref{30}) to first order of the GUP parameter become
\begin{equation}\label{37}
E=\frac{1}{2} \hbar  \omega  \coth \left(\frac{1}{2}b \hbar
\omega\right)+\frac{1}{4} \beta  m  \hbar ^{2}  \omega ^{2} \left[
\coth ^{2} \bigg( \frac{b  \hbar  \omega}{2} \bigg) - b \hbar \omega
\coth \bigg(\frac{b  \hbar  \omega}{2}\bigg) \sinh^{-2}\bigg
(\frac{b  \hbar  \omega}{2}\bigg) \right]+\mathcal{O}(\beta^{2}),  \end{equation}
\begin{equation}\label{38}
C=\frac{k_{B} \left(\frac{1}{2}b \hbar  \omega \right)^{2}}{\sinh
^{2} \left(\frac{1}{2}b  \hbar \omega\right)}+\frac{1}{2} \beta m
\hbar^{3}   \omega^{3}
 \left[ \dfrac{2 \big(\sinh(b \hbar \omega )- b \hbar \omega \big) - (b \hbar \omega ) \cosh(b \hbar \omega)}{  k_{B}  T^{2} \big ( \cosh(b \hbar \omega)-1 \big)^2}
 \right]+\mathcal{O}(\beta^{2}).
\end{equation}
 The condition $1 -b \, \varepsilon \, \frac{
\frac{
\partial ^{2} }{
\partial b^{2} } Z^{0} (b) }{ Z^{0}(b) } > 0$ in this case becomes
\begin{equation}\label{41}
b \bigg( \frac{1}{2} \, \beta \, m \bigg)  \bigg( \frac{\hbar^{2} \,
\omega^{2} }{4} \bigg) \bigg( -1+2 \coth ^{2} \bigg ( \frac{b  \hbar
\omega}{2} \bigg) \bigg) < 1,  \quad
\end{equation}
which for $\hbar\omega\ll k_BT $ we obtain $ \beta  m  k_{B} T < 1$.
Then the temperature condition (\ref{shart}) is given by
\begin{equation}\label{43}
\frac{\beta m}{2k_{B}} \ll T < \frac{1}{\beta m k_{B}}.
\end{equation}
Therefore, in the small $T$ limit, $E$ and $C$ read
\begin{eqnarray}\label{39}
  E&=&k_{B}  T-m  \beta  k^{2}_{B}  T^{2}+\mathcal{O}(\beta^2), \\
C&=&k_{B}-2m  \beta  k^{2}_{B}  T+\mathcal{O}(\beta^2).\label{40}
\end{eqnarray}
Again, the heat capacity tends to zero at $T= \frac{1}{2\beta m k_{B}}$ in agreement
with Ref.~\cite{27}. Similarly, for anti-Snyder GUP we have
\begin{eqnarray}\label{108}
\hspace{-1cm} E&=&\frac{1}{2} \hbar  \omega  \coth
\left(\frac{1}{2}b \hbar \omega\right)+\frac{1}{4} \beta  m \hbar
^{2}  \omega ^{2} \left[ \coth ^{2} \bigg( \frac{b  \hbar \omega}{2}
\bigg) - b \hbar
\omega  \coth \bigg(\frac{b  \hbar  \omega}{2}\bigg) \sinh^{-2}\bigg (\frac{b  \hbar  \omega}{2}\bigg) \right]+\mathcal{O}(\beta^{2}),\hspace{1cm} \\
\hspace{-1cm}C&=&\frac{k_{B} \left(\frac{1}{2}b \hbar  \omega
\right)^{2}}{\sinh ^{2} \left(\frac{1}{2}b  \hbar \omega
\right)}+\frac{1}{2} \beta  m \hbar^{3}   \omega^{3}
 \left[ \dfrac{2 \big(\sinh(b \hbar \omega )- b \hbar \omega \big) - (b \hbar \omega ) \cosh(b \hbar \omega)}{  k_{B}  T^{2} \big ( \cosh(b \hbar \omega)-1 \big)^2}
 \right]+\mathcal{O}(\beta^{2}).\label{109}
\end{eqnarray}
So in the small $T$ limit we find
\begin{eqnarray}\label{110}
E&=&k_{B}  T+m  \beta  k^{2}_{B}  T^{2}+\mathcal{O}(\beta^2), \\
C&=&k_{B}+2m  \beta  k^{2}_{B}  T+\mathcal{O}(\beta^2). \label{111}
\end{eqnarray}
We observe that the heat capacity does not tend to zero for this
case.

\subsection{The Landau levels}
Consider the problem of a particle of mass $ m $  and charge $ e $
in a constant magnetic field $ \vec{B}=B\hat{k} $, with potential $
\vec{A}=Bx\hat{j} $ in the Landau gauge. The expression for the
Hamiltonian is
\begin{equation}\label{44}
H^{0}= \frac{1}{2m} \big(\vec{p} -e\vec{A} \big),
\end{equation}
and the eigenfunctions and the eigenvalues of the Schr\"odinger
equation are \cite{9}
\begin{equation}\label{45}
\psi_{k,n}(x,y)=e^{iky}\phi_{n}(x-x_{0}),
\end{equation}
and
\begin{equation}\label{46}
E_{n}^{0}=\hbar \, \omega_{c}\Big(n+\frac{1}{2}\Big), \;  \quad n
\in N,
\end{equation}
where $ \omega_{c}=\frac{eB}{m} $ is the cyclotron frequency and $
\phi_{n} $ are the harmonic  oscillator eigenfunctions. In GUP
framework and to the first order of the GUP parameter, the
Hamiltonian of this system is given by  \cite{9}
\begin{equation}\label{47}
H=H^{0}+4 \beta m {H^{0}}^{2}.
\end{equation}
The eigenvalues are therefore
\begin{equation}\label{48}
E_{n}=E_{n}^{0}+4\beta m {E_{n}^{0}}^{2}
\end{equation}
and $\frac{\Delta E_{n}}{E_{n}^{0}}=4 \beta m \hbar \omega _{c}\Big
( n+\frac{1}{2}\Big)$. In ordinary quantum statistical mechanics,
the partition function of this system is obtained as
$Z^{0}(b)=\left[ 2\sinh \Big( \frac{b  \hbar  \omega _{c}}{2} \Big)
\right]^{-1}$. In analogous to harmonic oscillator, we obtain the
following expression for the internal  energy and the heat capacity
of this system and to first order of the GUP parameter as
\begin{eqnarray}\label{50}
\hspace{-1cm}E&=&E^{0}+\beta  m  \hbar^{2}  \omega_{c}^{2}\left[
-1+2 \coth ^{2} \Big( \frac{b  \hbar  \omega _{c}}{2} \Big) -2b
\hbar  \omega_{c}  \coth\Big (\frac{b  \hbar  \omega
_{c}}{2}\Big)\sinh^{-2}
\Big( \frac{b  \hbar  \omega _{c}}{2} \Big)  \right]+\mathcal{O}(\beta^{2}), \\
\hspace{-1cm} C&=&C^{0}+4\beta  m  \hbar^{3}   \omega_{c}^{3}
 \left[ \dfrac{2 \big(\sinh(b \hbar \omega_{c} )- b \hbar \omega_{c} \big) - (b \hbar \omega_{c} ) \cosh(b \hbar \omega_{c})}{  k_{B}  T^{2} \big ( \cosh(b \hbar \omega_{c})-1 \big)^2}
 \right]+\mathcal{O}(\beta^{2}) ,\label{51}
\end{eqnarray}
where  $E^{0}=\frac{1}{2} \hbar  \omega_{c}  \coth
\left(\frac{1}{2}b \hbar \omega_{c} \right)$ and $C^{0}=\frac{k_{B}
\left(\frac{1}{2}b \hbar  \omega_{c} \right)^{2}}{\sinh ^{2}
\left(\frac{1}{2}b  \hbar \omega_{c} \right)}$. Now, for
$\hbar\omega_c\ll k_BT  $ the temperature condition (\ref{shart})
reads
\begin{equation}\label{54}
\frac{4 \beta \, m}{k_{B} \, } \ll T < \frac{1}{\beta \, m \,
k_{B}}.
\end{equation}
For this case, the above relations in the small $T$ limit simplify to
\begin{eqnarray}\label{56}
 E&=&k_{B}  T - 8  \beta  m  k_{B}^{2}  T^{2}+\mathcal{O}(\beta^2),\\
 C&=&k_{B}-16\beta  m  k_{B}^{2}  T+\mathcal{O}(\beta^2).\label{57}
\end{eqnarray}
Note that although the energy spectrum depends on the magnetic field
$B$, the above equation shows that for $\hbar\omega_c\ll k_BT  $ the
internal energy is independent of the magnetic field. Again, for the
anti-Snyder GUP we have
\begin{eqnarray}\label{112}
\hspace{-1cm}E&=&E^{0}-\beta  m  \hbar^{2}  \omega_{c}^{2}\left[
-1+2 \coth ^{2} \Big( \frac{b  \hbar  \omega _{c}}{2} \Big) -2b
\hbar
\omega_{c}  \coth\Big (\frac{b  \hbar  \omega _{c}}{2}\Big)\sinh^{-2} \Big( \frac{b  \hbar  \omega _{c}}{2} \Big)  \right]+\mathcal{O}(\beta^{2}), \\
\hspace{-1cm} C&=&C^{0}-4\beta  m  \hbar^{3}   \omega_{c}^{3}
 \left[ \dfrac{2 \big(\sinh(b \hbar \omega_{c} )- b \hbar \omega_{c} \big) - (b \hbar \omega_{c} ) \cosh(b \hbar \omega_{c})}{  k_{B} T ^ {2} \big ( \cosh(b \hbar \omega_{c})-1 \big)^2}
 \right]+\mathcal{O}(\beta^{2}) ,  \label{113}
\end{eqnarray}
and
\begin{eqnarray}\label{115}
E&=&k_{B}  T + 8  \beta  m  k_{B}^{2}  T^{2}+\mathcal{O}(\beta^2), \\
 C&=&k_{B}+16\beta  m  k_{B}^{2}  T+\mathcal{O}(\beta^2), \label{116}
\end{eqnarray}
in the small $T$ limit.

\section{Conclusions}
In this paper, we have studied a GUP framework that implies a
nonzero minimal length uncertainty and presented an analytic method
to find the effects of GUP on the thermodynamical properties of the
physical systems. We applied the method on one-dimensional ideal
gas, harmonic oscillator, and Landau levels and obtained the
modified partition functions, internal energies, and the heat
capacities. We showed that up to the first order of the deformation
parameter, the shift in the internal energy and heat capacity are
quadratic and linear functions of the temperature, respectively. In
particular, the minimal length causes the heat capacity tends to
zero at some finite temperature in correspondence with the classical
result. We also studied the thermodynamics of these systems in the
anti-Snyder framework where the heat capacity represented a
non-vanishing behavior.

\end{document}